# Comment on Breakup Densities of Hot Nuclei

V.E. Viola[a], K. Kwiatkowski[b],

S.J. Yennello[c] and J.B. Natowitz[c]

[a]*IUCF and Department of Chemistry, Indiana University, Bloomington, IN 47405*

[b]*Los Alamos National Laboratory, Los Alamos, NM 87545.*

[c]*Department of Chemistry and Cyclotron Institute, Texas A & M University, College Station, TX 77843*

**Abstract**

In [1,2] the observed decrease in spectral peak energies of IMFs emitted from hot nuclei was interpreted in terms of a breakup density that decreased with increasing excitation energy. Subsequently, Raduta *et al.* [3] performed MMM simulations that showed decreasing spectral peaks could be obtained at constant density. In this letter we examine this apparent inconsistency.

## 1 Introduction

In a recent analysis of kinetic energy spectra for intermediate mass fragments (Z=2 < IMF < Z∼20) emitted in energetic light-ion-induced reactions, it was shown that the centroids of the Coulomb peaks systematically shift to lower energies and the widths broaden as a function of increasing excitation energy,

E*/A [1,2]. This result was interpreted as evidence for a decrease in the average breakup density from normal nuclear density $\rho_0$ at low excitation energies to a constant value of $\rho/\rho_0 \sim 0.3$ for E*A $\sim$ 4 MeV and above. Subsequently, Raduta *et al.* [3] performed MMM (microcanonical multifragmentation model) calculations [4] that showed that the observed centroid shifts could qualitatively be explained using a constant density of $\rho/\rho_0 \sim 0.2$. They concluded that " ... a decrease in the peak centroids for kinetic energy spectra can be observed at low constant density, which is different than published results in [1]". Here we examine this apparent conflict in interpretation.

First, it is relevant to clarify the method used to fit the experimental spectra from which the experimental breakup densities were derived. The parameterization of the spectral data was based on the model of Moretto [5], as adapted by Kwiatkowski [6]. The model extends the fission transition state model of Nix and Swiatecki [7] to describe all mass divisions, accounting for the evolution of spectral shapes from Gaussian for fission to Maxwellian for nucleon emission. In addition to containing Coulomb barrier, slope(temperature) and kinematic parameters, the model has the important advantage of allowing for a broadening of the spectral widths with increasing excitation energy to account for fluctuations that accompany higher E*/A values, as has long been known for fission [8].

For present purposes the primary concern is the Coulomb parameter, which depends on the average charge separation distance between a given fragment and the charge distribution of the remaining system. Using fission fragment kinetic-energy systematics [9] as a reference point for the charge separation distance at low excitation energies, the parameterization in [1,2] provides good fits to IMF spectra at E*/A values below 2 MeV, for which $\rho/\rho_0 \sim 1$. For higher excitation energies, satisfactory fits can only be obtained with a reduced Coulomb barrier parameter, which implies a larger separation distance for the breakup configuration and hence a lower density. In the excitation-energy regime where binary breakup is the dominant de-excitation mechanism, the parameterization of [1,2] should provide a satisfactory description of the IMF-heavy residue charge separation distance at breakup. For multifragmentation events, it is assumed that this model provides a first-order approximation to the average Coulomb field experienced by individual IMFs at breakup.

The top frame of Fig. 1 shows the average Coulomb parameters derived from the moving source fits in [1,2]. The fits take into account the significant decrease in source size with increasing E*/A, which coupled with the decreasing spectral peak centroids, leads to a near constant behavior at high excitation energies. For example, at E*/A = 8 MeV the source is about 75% of the target mass due to fast cascade and preequilibrium processes. Shown in the bottom frame of Fig. 1 are the extracted experimental densities, which assume a spherical breakup geometry. Both the Coulomb parameter and the densities are observed to decrease rapidly in the range E*/A = 1 - 4 MeV, after which they become constant within experimental error. The upshot is that up to E*/A $\sim$ 4 MeV the analysis shows a decreasing density as a function of increasing excitation energy - which disagrees with the MMM calculations. However, above E*/A $\sim$ 4 MeV the decreasing Coulomb peak centroids lead to a nearly constant density, which is the same conclusion the Raduta *et al.* predict.

In order to place the experimental results and the MMM simulations in perspective, it is useful to compare the IMF multiplicity distributions, as shown in Table 1. The experimental distributions for specific multiplicities are shown in

Table 1

Comparison of IMF multiplicities and breakup densities between the experimental values of [1,2] and MMM simulations [3].

| E*/A (MeV) | 1.0 | 2.0 | 3.4 | 4.5 | 5.7 | 6.8 | 7.9 | 10.0 |
|---|---|---|---|---|---|---|---|---|
| $M_{IMF}$ | 0.06 | 0.18 | 0.5 | 1.7 | 2.6 | 3.3 | 4.0 | 4.7 |
| $M_{IMF}$(MMM) | * | * | 2.8 | 4.5 | 6.0 | 7.6 | 8.9 | 10.1 |
| M(exp)/M(MMM) | * | * | 5.6 | 2.6 | 2.3 | 2.3 | 2.3 | 2.1 |
| $<\rho/\rho_0>$(exp) | 1.0 | 0.95 | 0.6 | 0.41 | 0.26 | 0.32 | 0.31 | * |
| $\rho/\rho_0$(MMM) | * | * | 0.2 | 0.2 | 0.2 | 0.2 | 0.2 | 0.2 |

Fig. 2 , along with the probabilities for events with $M_{IMF} > 2$. The multiplicity data show that below E*/A $\sim$ 4 MeV, IMF emission is primarily a binary process. For higher excitation energies, the multiplicities increase rapidly, signaling the onset of multifragmentation as the dominant decay mechanism. With respect to the present discussion, the salient point is that multifragmentation does not become a significant decay process until the excitation energy has exceeded this threshold and therefore comparison with statistical multifragmentation models such as MMM should be applied with caution at lower energies.

The comparison of the absolute values and the multiplicity ratios in Table I reveals significant discrepancies between the data and the MMM simulations. The IMF multiplicity and the charge distributions are basic observables that one expects to be reproduced with any multifragmentation model. Two observations are apparent in the table. First, for E*/A below 4 MeV, the MMM model increasingly overpredicts the multiplicities as the excitation energy de-creases. Second, above this value the model/data ratio is a constant value of about 2.3. By increasing the breakup density in the model, this discrepancy could probably be resolved. Thus, it is useful to separate the analysis into two excitation energy regimes, above and below $E^*/A \cong 4$ MeV.

First, we examine the higher $E^*/A$ results. In this regime, the experimental results of [1,2] lead to the same conclusion as the MMM model; i.e., *the kinetic energy centroids shift systematically to lower energies with increasing excitation energy at constant density*. This result has been shown previously with the simultaneous SMM [10] model in [11,12], a calculation that also reproduces the experimental charge distributions and the multiplicities. Thus, in the regime where multifragmentation is the dominant decay mode, the simulations presented by Raduta et al, which use a more sophisticated Coulomb calculation for multifragment events, reinforce previous conclusions. In order for the MMM model to yield more quantitative agreement in this region, its parameters need to be modified to reduce the multiplicities by a factor of two and to reproduce the charge distributions. In addition, the calculation needs to include the significant decrease in source size at higher excitation energy.

Irrespective of any parameterization, the observed centroid decrease with increasing $E^*/A$ must logically be explained at least in part by a decrease in the density, as originally pointed out by Poskanzer [13]. The sequential EES model [14] and the metastable mononucleus model [15] have been shown to describe the density evolution in this low energy regime. Raduta *et al* do not address this necessary evolution in the breakup density from $\rho/\rho_0 = 0.2$ assumed in their model to $\rho/\rho_0 = 1.0$ at low excitation energy.

In conclusion, for excitation energies where multifragmentation is the domi-nant breakup mechanism, i.e. E*/A > 4 MeV, the MMM simulations of [3] and the experimental results [1,2] both show a decrease in the IMF spectral peaks with increasing E*/A at constant density. *Thus, in this regime, the experimental results and the calculations are qualitatively self-consistent.* For lower excitation energies, binary breakup is the dominant decay mode and analysis of the spectra with a multifragmentation mechanism cannot account for the data.

Acknowledgements: This work was supported by the U.S. Department of Energy.

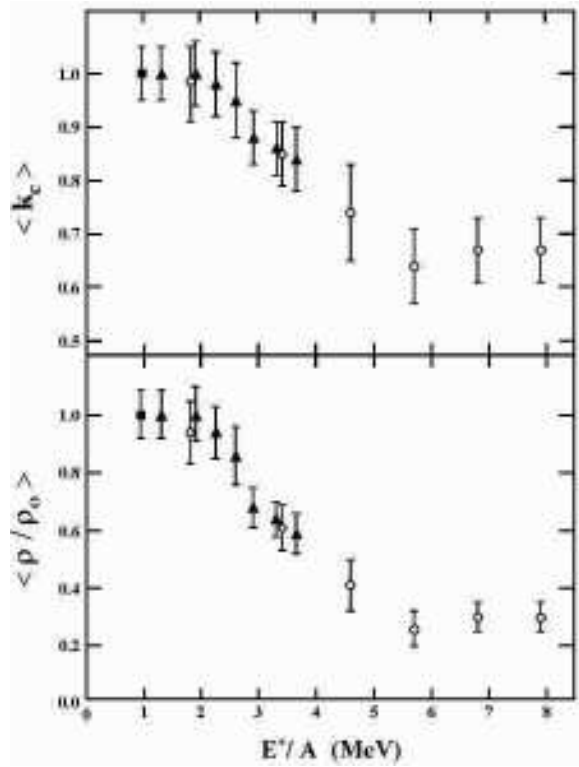

Fig. 1. Top: Dependence of the average moving-source Coulomb parameter $<k_c>$ as a function of excitation energy. Symbols are as follows: 200 MeV $^4$He ($\square$); E/A = 20-100 MeV $^{14}$N ($\triangle$); 4.8 GeV $^3$He ($\square$). Bottom: Average density $<\rho/\rho_0>$ as a function of E*/A derived from the $k_c$ values in the top panel.

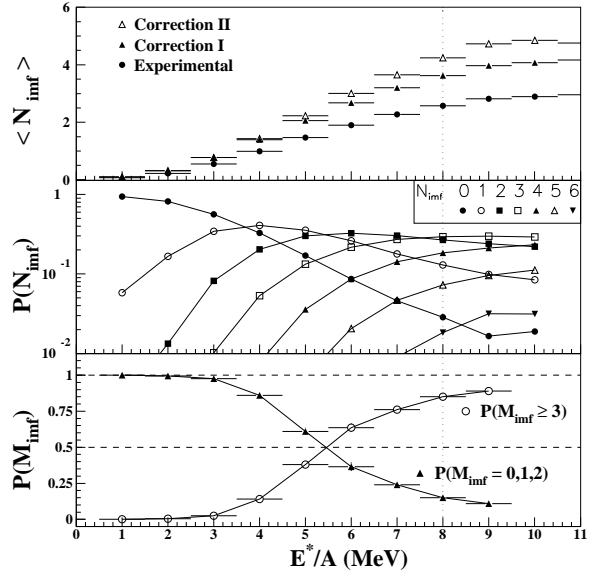

Fig. 2. Top: Average number of IMFs as a function of excitation energy for the 8 GeV/c $\pi^- + {}^{197}$Au reaction: observed $N_{imf}$ (circles), corrected for geometry (solid triangles) and corrected for both geometry and fragment energy thresholds $M_{imf}$ (open triangles. Middle: probability for a given number of observed IMFs. Bottom: Probability for corrected IMF multiplicity $M_{imf} > 2$ (circles) and $M_{imf} < 3$ (solid circles).